\input{aipcheck}

\documentclass[
    ,final            
  ]
  {aipproc}

\layoutstyle{6x9}

\begin{document}

\title{High Resolution Spectroscopy of SN1987A's Rings: He I
  $\lambda$10830 and H$\alpha$ from the Hotspots}


\author{Nathan Smith}{
  address={Astronomy Department, University of California, 601
  Campbell Hall, Berkeley CA 94720} }
\author{Richard McCray}{
  address={JILA, University of Colorado, 440 UCB, Boulder, CO 80309}
}

\begin{abstract}

We present the first high-dispersion spectroscopy of He~{\sc i}
$\lambda$10830 from the hotspots in the ring around SN1987A, obtained
at Gemini South, spatially resolving the near and far sides of the
ring.  We compare these line profiles to similar echelle spectra of
H$\alpha$ and [N~{\sc ii}] $\lambda$6583 obtained at the Magellan
Observatory.  We find that the He~{\sc i} profiles are much broader
than H$\alpha$ or [N~{\sc ii}], but the He~{\sc i} profiles also have
different shapes -- they have enhanced emission at high speeds, with
extra blueshifted emission on the north side of the ring, and extra
redshifted emission on the south side.  To explain this, we invoke a
simple geometric picture where the extra He~{\sc i} emission traces
hotter gas from faster shocks that strike the apex of the hotspots
directly, while the H$\alpha$ preferentially traces cooler
lower-ionization gas from slower transverse shocks that penetrate into
the sides of the ring.

\end{abstract}
\maketitle

\section{INTRODUCTION}

Twenty years after the explosion, the blast wave from SN1987A has now
reached and is plowing through the circumstellar ring.  Protrusions
caused by Rayleigh-Taylor instabilities in the ring were hit first by
the blast wave, giving rise to a series of ``hotspots'' around the
ring [2,5,6].
The interaction has three different velocity components:

1.  Extremely broad ($-$15,000 to 15,000 km s$^{-1}$) emission in
    H$\alpha$ and Ly$\alpha$ that traces H atoms crossing the reverse
    shock.  This can be seen in low-resolution spectra [4].  One sees
    blueshifted emission to the north, and redshifted emission to the
    south, like the expansion pattern of the ring itself [1].

2.  Broad (few 10$^2$ km s$^{-1}$) components emitted by gas in the
    ring that has been passed over by the forward shock [2,3].  This
    is the emission from the ``hotspots''.

3.  Narrow (10's of km s$^{-1}$) emission lines from slow-moving gas
    in the circumstellar ring that has not yet been reached by the
    shock, but is photoionized by UV emission from the shock
    interaction [3].

Here we concentrate on the spatially resolved emission-line profiles
of the shock-heated has in the hotspots on the north (near) and south
(far) sides of the ring, seen in H$\alpha$ and He~{\sc i}
$\lambda$10830.

\begin{figure}
  \includegraphics[height=.5\textheight]{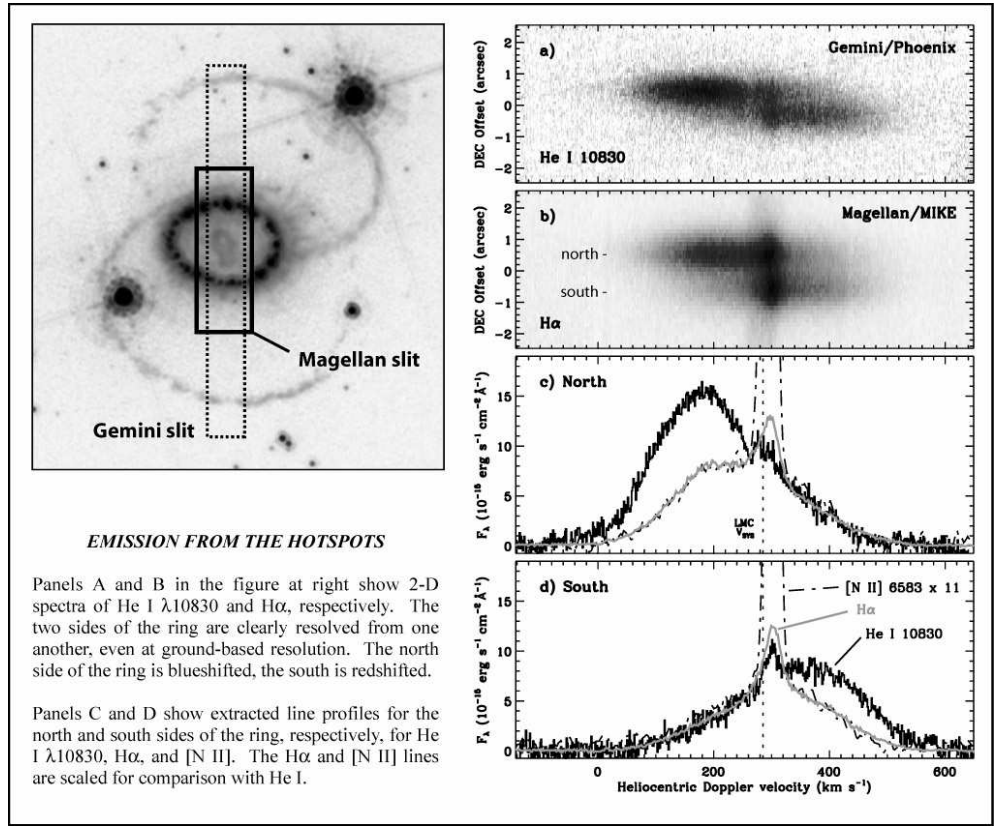}
  \caption{He~{\sc i}, H$\alpha$, and [N~{\sc ii}] spectra.}
\end{figure}


{\bf Observations:} We present high resolution (R=60,000) ground-based
spectra of the inner equatorial ring of SN~1987A.  We used the Phoenix
spectrograph on Gemini South to observe He~{\sc i} $\lambda$10830 in
Apr 2006, with the 0.5'' slit oriented as in Fig.\ 1.  These long-slit
data spatially resolved the north and south sides of the ring (Fig.\
1a).  We also used the MIKE echelle spectrograph at the Magellan
Observatory to obtain the optical spectrum in March 2005.  The 2-D
spectrum of H$\alpha$ is shown in Fig.\ 1b.  Tracings of the north and
south sides of the rings are shown in Figures 1c and d, respectively,
for both lines as well as [N~{\sc ii}] $\lambda$6583.  The [N~{\sc
ii}] line has stronger narrow emission from unshocked circumstellar
gas, and is scaled up to show that the profile of its broad shocked
component is identical to H$\alpha$.

\begin{figure}
  \includegraphics[height=.38\textheight]{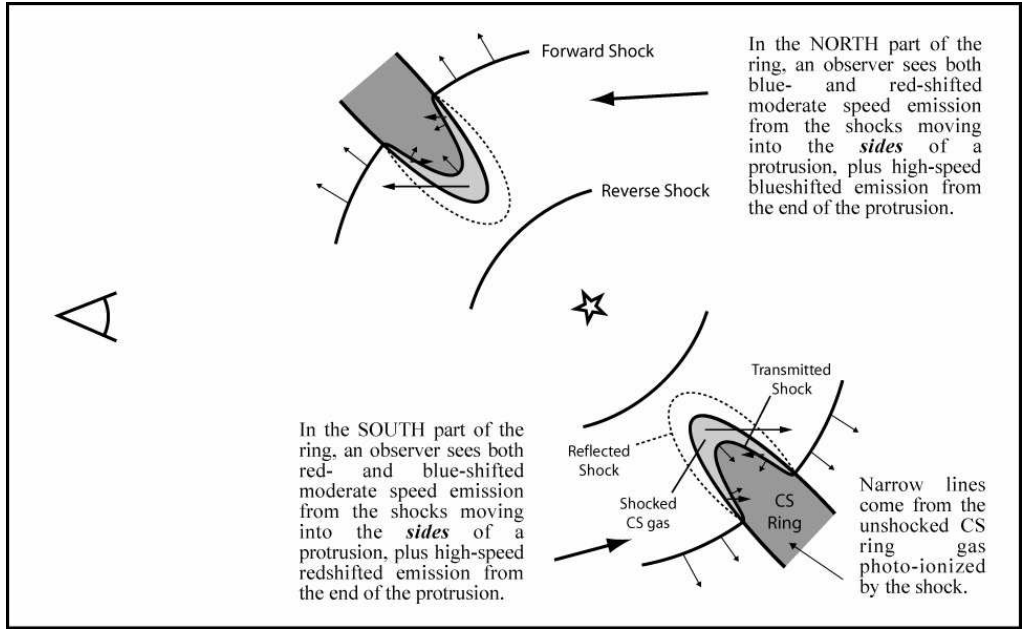}
  \caption{Sketch of the geometry that may cause the different He~{\sc
  i} and H$\alpha$ line profiles in Fig.\ 1.}
\end{figure}

\section{LINE PROFILES AND GEOMETRY OF HOTSPOT EMISSION}

In the scenario depicted in Figure 2, we can understand the difference
in line profile shape between He~{\sc i} $\lambda$10830 and H$\alpha$
as a consequence of ionization levels, geometry, and shock kinematics.
When the expanding blast wave encounters an obstacle like an
equatorial ring, with or without a protruding ``finger'' from
Rayleigh-Taylor instabilities, the shock will penetrate the dense
material head-on and decelerate, but it will also wrap around the
object, transmitting weaker and slower oblique shocks into the sides
of the obstacle (see Fig.\ 2).  The denser, cooler, slower, and
lower-ionization gas in these oblique shocks probably dominates the
H$\alpha$ and broad [N~{\sc ii}] emission [3].  This will produce
broad emission at both blue and redshifted velocities.  He~{\sc i}
$\lambda$10830 emission will arise in this same gas, but it is also
likely to be enhanced relative to H$\alpha$ in the hotter and higher
ionization gas at the head of the shocked column.  If true, that would
in principle explain the observed line profiles, because that gas would
have higher projected speeds on both the blue and red sides of the
ring (Fig. 2) than the circumstellar gas struck by the oblique shocks
in the side of the column, as observed (Fig.\ 1).

The enhanced He~{\sc i} emission at the end of the shocked column,
plus the weakness of the narrow He~{\sc i} emission from unshocked gas
(less than 5\% of the total) means that He~{\sc i} $\lambda$10830
images of the hotspots [6] are among the best tracers of the strong
shocks in the circumstellar gas.  A comparison of variability between
H$\alpha$ and He~{\sc i} in high-resolution images might provide
important clues about the geometry of the shock front.  If the
``fingers'' that are being illuminated by the shock at the present
time to produce the hotspots really are the inner protrusions from a
more massive ring, then we should expect the ring to brighten
dramatically in the near future in He~{\sc i} $\lambda$10830 as this
more extended material is overtaken by the main blast wave.




\begin{thebibliography}{9}
%

\bibitem{ch00}
A.P.S.\ Crotts, and S.R.\ Heathcote,  \emph{ApJ}, \textbf{528},
426--435 (2000)

\bibitem{michael00}
E.\ Michael, et al., \emph{ApJ}, \textbf{542}, L53--L56 (2000)

\bibitem{pun02}
C.S.J.\ Pun, et al., \emph{ApJ}, \textbf{572}, 906--931 (2002)

\bibitem{smith05}
N.\ Smith, S.\ Zhekov, K.\ Heng, R.\ McCray, J.A.\ Morse, M.\
Gladders, \emph{ApJ}, \textbf{635}, L41--L44 (2005)

\bibitem{sonn98}
G.\ Sonneborn, et al., \emph{ApJ}, \textbf{492}, L139-L142 (1998)

\bibitem{sugar}
B.E.K.\ Sugerman, et al., \emph{ApJ}, \textbf{572}, 209--226 (2002)

\end{thebibliography}
\end{document}